# Ultra-low transmission loss (7.7 dB/km at 750 nm) inhibited-coupling guiding hollow-core photonic crystal fibers with a single ring of tubular lattice cladding

B. Debord,[1,2] A. Amsanpally,[1] M. Chafer,[1,2] A. Baz,[1] M. Maurel,[1,2] J.M Blondy,[1] E. Hugonnot,[3] F. Scol,[3] L. Vincetti,[4] F. Gérôme,[1,2] and F. Benabid[1,2*]

[1] *GPPMM Group, Xlim Research Institute, CNRS UMR 7252, University of Limoges, France.*
[2] *GLOphotonics S.A.S, 1 avenue d'Ester, Ester Technopôle, Limoges, France.*
[3] *Commissariat à l'Energie Atomique et aux Energies Alternatives, Centre d'Etudes Scientifiques et Techniques d'Aquitaine, 33116 Le Barp Cedex, France.*
[4] *Department of Engineering "Enzo Ferrari", University of Modena and Reggio Emilia, I-41125 Modena, Italy.*

*Corresponding author: benoit.debord@xlim.fr and f.benabid@xlim.fr*

**The advent of photonic bandgap (PBG) guiding hollow-core photonic crystal fiber (HC-PCF) sparked the hope of guiding light with attenuation below the fundamental silica Rayleigh scattering limit (SRSL) of conventional step-index fibers. Unfortunately, the combination of the strong core-cladding optical-overlap, the surface roughness at the silica cladding struts and the presence of interface-modes limited the lowest reported transmission-loss to 1.2 dB/km at 1550 nm. This hope is recently revived by the introduction of hypocycloid core-contour (*i.e.* negative curvature) in inhibited-coupling (IC) guiding HC-PCF, and the reduction of their confinement loss to a level that makes them serious contenders for light transmission below the SRSL in UV-VIS-NIR spectral range. Here, we report on several IC guiding HC-PCFs with a hypocycloid core-contour and a cladding structure made of a single ring from a tubular lattice. The fibers guide in the UV-VIS and NIR, and among which we list one with a record transmission loss of 7.7 dB/km at ~750 nm (only a factor ~2 above the SRSL), and a second with an ultra-broad fundamental-band with loss in the range of 10-20 dB/km spanning from 600 to 1200 nm. Both fibers present near-single mode guidance and very low bend loss sensitivity. The results show that the limit in the transmission is set by confinement loss for wavelengths longer than ~1 µm and by surface-roughness for shorter wavelengths, thus indicating that transmission loss well below the SRSL in the visible is possible with a surface roughness reduction and would open the possibility of the first UV low-loss light-guidance.**

## 1. Introduction

Since its first proposal [1], hollow-core photonic crystal fiber (HC-PCF) has proved to be a remarkable platform for our understanding on how light is confined in such microstructured waveguides, and an outstanding host for gas-laser applications [2], [3].

In HC-PCF, a core-guided mode is prevented from coupling to the cladding by either Photonic Bandgap (PBG) [1] or Inhibited-Coupling (IC) [4]. In PBG guiding fibers, the coupling of the core mode to the cladding is disallowed by micro-engineering the cladding structure such its modal spectrum is void from any propagation modes at the core guided-mode effective index-frequency space, $n_{eff} - \omega$ (*i.e.* $\{|\varphi_{clad}\rangle\} = \emptyset$, with $\varphi_{clad}$ being the field wave function of the propagating mode in the cladding structure) [1]. It is this absence of cladding modes that restraints a core mode from leaking out through the cladding and forces to be guided in a PBG fiber-core. For the case of IC guidance, the condition of a PBG is no longer required. In analogy with quasi-bound and bound state in a continuum (QB-BIC), first proposed by Van Neumann and Wigner in 1929 within the context of quantum mechanics [5], a core guided-mode and cladding modes can have the same effective index, and yet propagate without strongly interacting. Though QB-BIC remains uncommon scheme of trapping waves, it is universal [6], and has been reported in photonic crystal [7], electronic [8] and mechanical [9] systems. In optical fiber form, QB-BIC works as follows: a core guided-mode (often leaky mode) cannot, or strongly inhibited to, channel out through the cladding by a strong reduction in the coupling between the cladding modes and a core mode. In other word, the dot product between the field of the core mode $|\varphi_{core}\rangle$ and the cladding mode $|\varphi_{clad}\rangle$ is strongly reduced (*i.e.* $\langle\varphi_{clad}|\Delta n^2|\varphi_{core}\rangle \rightarrow 0$, with $\Delta n$ being the photonic structure transverse index profile). This core-cladding coupling factor $\langle\varphi_{clad}|\Delta n^2|\varphi_{core}\rangle$ can be reduced by either having little spatial intersection between the fields of $|\varphi_{clad}\rangle$ and $|\varphi_{core}\rangle$ or by having a strong mismatch in their respective transverse spatial phase. Drawing on ideas from solid-state physics, the light leakage is reduced in IC fibers owing to symmetry incompatibility between the transverse phase of the core mode and those of the cladding modes [6]. In this situation, core guidance can occur even with a dense continuum of cladding modes at its $n_{eff} - \omega$ space provided that the latter have a strong transverse-phase mismatch with the core guided-mode, and their intensity is concentrated in a cladding region with different material-index than that of the core [4]. In the sense that a PBG fiber is defined by the absence of phase-matched cladding modes to transfer light outward from the core, the cladding continuum in IC guiding fibers effectively acts as a quasi-PBG over the core mode $n_{eff} - \omega$ space.

In a previous paper [4], we laid down the design parameters for optimal IC guidance in air-silica HC-PCF such as Kagome lattice cladding. We have shown that IC guidance is enhanced by having a glass micro-structure exhibiting a very thin and elongated glass-membrane network with a minimum number of sharp bends or nodes, and by operating in the large pitch regime or high normalized frequency (*i.e.* the lattice pitch is larger than the optical wavelength $\lambda$) [10]. These structural features ensure modes with strong localization in the thin silica-web and with a very fast transverse field-oscillation (*i.e.* high azimuthal-like number). Furthermore, unlike with PBG, the criterion of $\langle\varphi_{clad}|\Delta n^2|\varphi_{core}\rangle \rightarrow 0$, which is directly linked to the reduction of the fiber confinement loss (CL) in IC guidance implies a strong dependence on the core-mode profile, and thus the core-contour. This led to the seminal introduction of hypocycloid core-contour design (*i.e.* negative curvature) in 2010 [11], [12], and which resulted in a dramatic reduction in transmission loss, as illustrated in the 17 dB/km at 1 µm achieved with Kagome-lattice HC-PCF and its record optical power handling [13][14]. This in turn has triggered a wide effort in designing hollow-core fibers with negative curvature [15], [16]. In a hypocycloid core-contour fiber, the core has a contour exhibiting a set of alternating negative curvature cups with inner radius $R_{in}$, and supporting an $HE_{11}$ guided-mode whose cross-section is identical to that of capillary with a radius equal to $R_{in}$. Consequently, the field overlap integral of the core-mode and the highly oscillating silica core-surround mode (cladding-mode) is strongly reduced compared to the previous circular or hexagonal contours in Kagome HC-PCF [10] via three origins. Firstly, the $HE_{11}$ spatial overlap with silica at $R_{in}$ is reduced to the tangent sections of the inner cups of the hypocycloid contour. Secondly, the larger perimeter of the hypocycloid contour compared to circle with radius $R_{in}$, results in a higher azimuthal-like number in the silica core-surround modes and hence, stronger transverse phase-mismatch with the core mode. Finally, the spatial overlap between the core-mode with connecting nodes -which support low azimuthal number modes- is strongly reduced [12], [17], [18].

The above features for IC guidance are also fulfilled in a tubular lattice, which consists of an arrangement of isolated thin glass tubes [19]. Such a cladding structure not only can exhibit fast transverse spatial phase thanks to its circular architecture but its fiber form exhibits a hypocycloid core-contour [15]. More crucially, when the cladding is reduced to a single ring, one can fabricate a fiber with a core-contour that is completely void of any connecting nodes, which strongly favors IC. As a matter of fact, because of the very weak $HE_{11}$ core-mode field overlap with the cladding in IC guiding fibers, low confinement loss can be expected with only one cladding ring. This is experimentally illustrated by the results obtained with a hollow-core fiber having a single-ring of this cladding lattice [15], [16]. In these works, the fiber cladding tubes were relatively thick (>1 µm), and the achieved transmission loss was as low as 34 dB/km in 3-4 µm spectral range. Another recent work achieved 30 dB/km at 1 µm using tube thickness of 830 nm [20].

Here, by optimizing such a tubular cladding and operating with much thinner glass-tube than has previously been achieved, we report on the fabrication of several ultra-low loss single-ring tubular lattice HC-PCFs (SR-TL HC-PCF) guiding in the UV-VIS and NIR. Among the fabricated fibers, we count one with a record transmission loss of 7.7 dB/km at ~750 nm, which is only 4 dB above the Rayleigh scattering fundamental limit in silica, and observed guidance down to 220 nm. A second one exhibits ultra-broad fundamental band with loss range of 10-20 dB/km over one octave spanning from 600 to 1200 nm. Both fibers present a near-single modedness, with ~20 dB extinction between $HE_{11}$ fundamental core mode and the strongest first higher-order mode (HOM), and exhibit very low bend loss (0.03 dB/turn for a 30 cm bend diameter at 750 nm). Finally, the comparison of the measured attenuation loss with theory shows that the fibers optical transmissions are limited by confinement loss (CL) for wavelengths longer than 1 µm (*i.e.* design-limited transmission performance) and by surface-roughness scattering loss (SSL) for shorter wavelengths. The surface roughness induces both SSL via scattering off the core-contour surface, and an increase in CL via stronger core-cladding mode coupling due to the higher sensitivity to index-fluctuations of the modal structure of the tubes at shorter wavelengths.

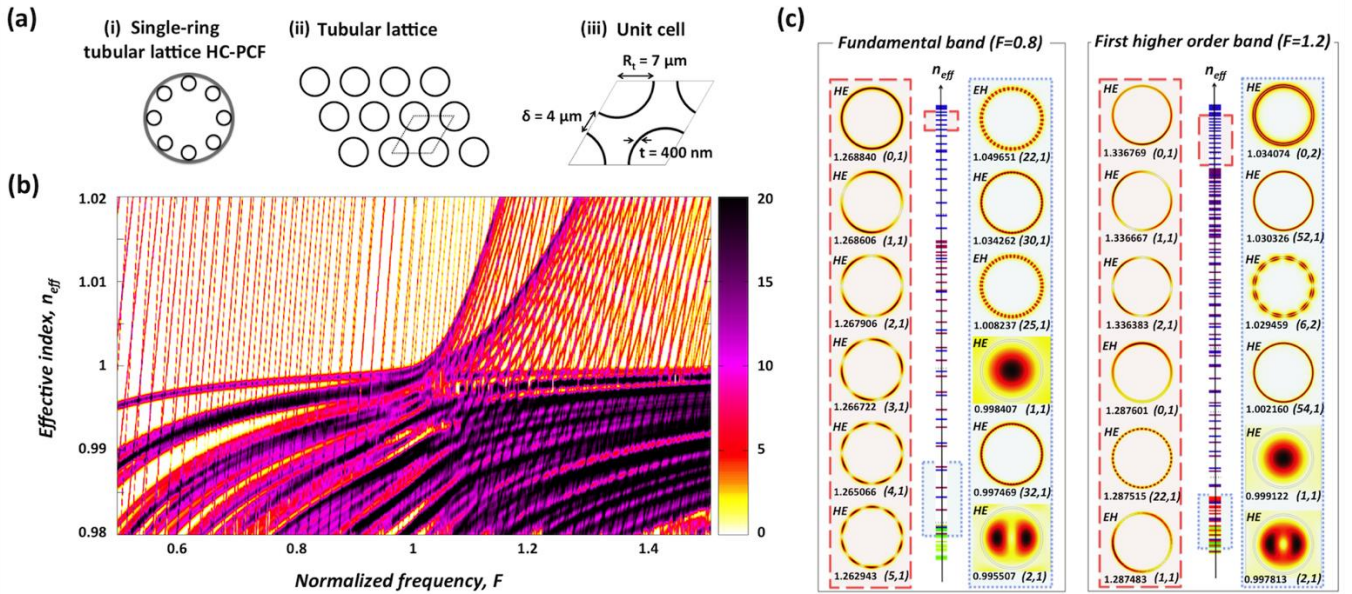

Fig. 1. (a) Schematic of HC-PCF with a single-ring tubular lattice (i), a section of a tubular lattice in a triangular arrangement (ii) and the details of a unit cell of the lattice (iii). (b) The density of the photonic state of an infinite tubular lattice in triangular arrangement. (c) Transverse profiles of the electric field magnitude of the photonic state (modes) for F = 0.8 (fundamental band) and for F = 1.2 (first higher order band).

## 2. Fiber parameter design

The left panel of Fig. 1(a) shows an example of the SR-TL HC-PCF that we are exploring. To understand and determine the design parameters for fiber optimal guidance, we first examine the nature of its cladding modes. For this purpose, one can consider the SR-TL HC-PCF cladding as a section from an infinite lattice of isolated tubes and where the fiber-core is introduced as a defect within the lattice. Furthermore, we operate in the large pitch regime, and thus the spectral characteristics of the photonic structure is primarily governed by the modal properties of the lattice single unit rather than their arrangement in the lattice [21]. This is explained using the photonic tight binding model [10] where for a lattice pitch much larger than the wavelength, the modes of each tube in its unit cell don't strongly interact with their close neighbors. For example, the modal spectrum of a cladding with $M$ tubes is comprised by $M$-degenerate modes from the individual tubes. Consequently, only the thickness and index of the unit cell individual waveguiding features (here silica tubes and air-holes) are relevant for the modal spectrum of the cladding photonic structure. In other word, under this large pitch regime the modal spectrum of the SR-TL HC-PCF cladding can be determined by considering an infinite lattice of any symmetry arrangement. We have confirmed this by numerically calculating the DOPS of tubular lattice for both triangular and square arrangement and for different lattice spacing, and found that the structure and content of modal spectrum to depend very little on the tube arrangement or spacing (see Supplement 1) [22]. Here, we take a triangular arrangement to calculate the modal spectrum of the SR-TL HC-PCF cladding (see central and right panel of Fig 1(a)).

Figure 1(b) maps the photonic structure propagation modes in the space $n_{eff} - F$ for a lattice of isolated glass tubes of refractive index $n_g = 1.45$ and with representative radius $R_t = 7~\mu m$, thickness $t = 400~nm$ and gap between two adjacent tubes $\delta = 4~\mu m$ (corresponding to a lattice pitch of 18 µm, which is much larger than any of our operating wavelengths). Here, $F = (2t/\lambda)\sqrt{n_g^2 - 1}$ is a normalized frequency relative to the tube thickness and the refractive index of the glass and air. The plot shows the density of photonic states (DOPS), defined to be the number of Maxwell equation mode solutions of the lattice at the normalized frequency $F$ in the range of effective index between $n_{eff}$ and $n_{eff} + \delta n_{eff}$, with $\delta n_{eff}$ chosen to be $10^{-5}$.

The DOPS was computed by using the commercial finite element modal solver COMSOL Multiphysics and by applying the Floquet-Bloch boundary conditions to the unit cell of the lattice. Here, we mapped the three symmetry points of the Brillouin zone.

As expected, the DOPS diagram shows no PBG regions in the mapped $n_{eff} - F$ space, but a quasi-continuum of photonic states. Within the explored $F$ spectral range of 0.5-1.5, we can distinguish two bands at each side of $F = 1$. For $F<1$, corresponding to the fiber transmission fundamental band, the modal spectrum exhibits a lower DOPS compared to the range $F>1$, which corresponds to the fiber 1st higher-order band. The color-map of these states highlights two classes of modes. The modes residing in the silica tube, which are identifiable with their steep dispersion curves (close to vertical), and those residing in air having flat dispersion curves (close to horizontal) and located below the vacuum-line (i.e. $n_{eff} = 1$). The nature of these modes is revealed in Fig. 1(c) by showing the transverse profile of their field magnitude for a given frequency from the fundamental band ($F = 0.8$) and from the 1st higher-order band ($F = 1.2$) at two representative sets of $n_{eff}$. The first set corresponds to $n_{eff} > 1$ and ranges from 1.2 to 1.3 (see the shaded light-red rectangle of Fig. 1(c)). Within this effective index range, the modes are all located in the silica tube as expected, and correspond to HE$_{ml}$ (i.e. electric field direction is azimuthal) or EH$_{ml}$ (i.e. electric field direction is radial) [23]. Here, $m$ and $l$ are the indices of the azimuthal and radial components of the transverse wavenumber respectively. The azimuthal and radial phases are commonly inseparable, however, given the fact that the radius of the tube is much larger than its thickness we can approximate the radial phase of these modes to that of a slab and apply the self-consistency condition on the azimuthal direction [23]. Consequently, for a fixed $\lambda$ and for $n_{eff}$ equal to that of HE$_{11}$ fiber core mode (i.e. the silica cladding modes are longitudinally phase-matched with the core mode), we can relate the azimuthal and radial numbers to $R_t$, $t$ and the fiber core radius $R_c$ through the identity:

$$\left(\frac{m+1/2}{2R_t}\right)^2 + \left(\pi\frac{l+1/2}{t}\right)^2 = \left(\frac{2.405}{R_c}\right)^2 \tag{1}$$

From this simple identity we can relate $m$ to $l$ and extract scaling laws such as $m \propto R_t/R_c$ when we fixed $t$ and the wavelength.

By definition, all the silica modes in the fundamental band don't exhibit radial oscillations (i.e. $l=1$), and EH$_{m,1}$ can be distinguished from HE$_{m,1}$ by their slight lower dispersion slope in the DOPS diagram. In the 1st higher order band, the radial number *l* can take either the value of 1 or 2, and the modes with *l*=2 can be distinguished by exhibiting two radial lobes in their field amplitude profile. The second set of $n_{eff}$ corresponds to the modes below the vacuum-line ($n_{eff} < 1$), and ranges from 1.04 and 0.99 (see the dotted blue rectangle of Fig. 1(c)). Below the vacuum line, the photonic structure can support modes in silica and air. Here, the modal content comprises both the above silica tube modes and those residing in the air of the tube inner-region (air modes). We note that because *m* increases as $n_{eff}$ decreases, all silica modes exhibit larger azimuthal number than those from the first effective-index set, which strongly favors coupling inhibition via enhanced transverse phase-mismatch between these silica modes and the low azimuthal number modes residing in air. This fact is exemplified by the easily distinguishable dispersion curve of the air modes in the DOPS. Indeed, the DOPS diagram shows that, apart from for frequencies near $F = 1$, where the dispersion of the air modes and the silica modes strongly hybridize (anti-cross), the dispersion of the two classes of modes are clearly distinguishable and they intersect with no meaningful anti-crossing, indicative of a strong coupling inhibition between them. The strong coupling between the air modes and the silica modes is practically limited to frequencies with integer values of $F$ (i.e. $F = 1,2,3,...$), which stems from a radial transverse phase-matching between silica and air modes when the azimuthal is nill or very low [19]. This air-glass modal coupling is manifested in the hollow-core fiber form of the photonic structure as high loss bands centered at wavelengths given by $\lambda_l = (2\,t/l)\sqrt{n_g^2 - 1}$. On a side note, we highlight that this radial (here *m*=0) transverse-phase matching condition is used in some of the current literature as a defining feature of fibers guiding via anti-resonance reflecting optical waveguide (ARROW) [21], [24]. We argue that the ARROW picture doesn't give the physical mechanisms by which the core guided modes is prevented from coupling to the cladding modes. In Supplement 1, we detail this point and provide some distinctive properties to differentiate between PBG, IC and ARROW, and on why the latter is not appropriate in comprehensibly describing the guidance in fibers such as SR-TL HC-PCF.

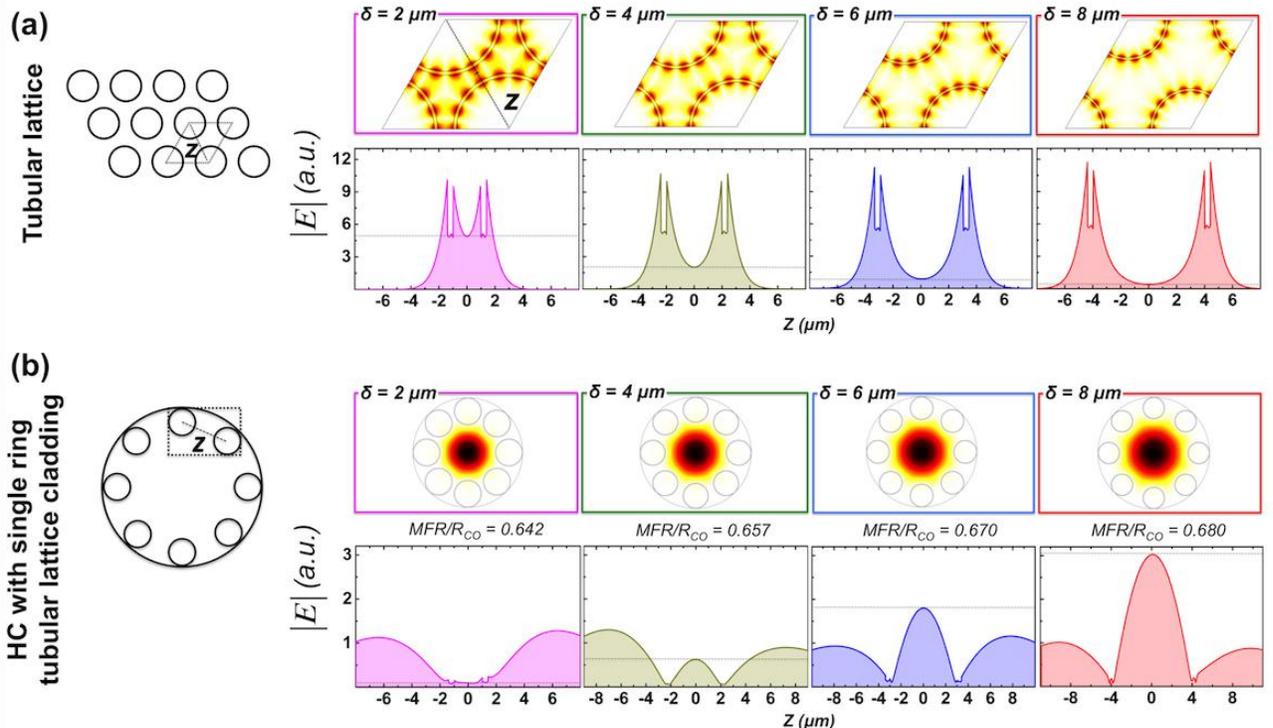

Fig. 2. 1D field amplitude profile along a gap line between the center of two adjacent tubes for spacing varying from 2 to 8 µm for a tubular lattice (a) and for hollow-core with single ring tubular lattice cladding (b).

Here, we use the IC formalism as a design tool to determine $R_t$, $t$ and $\delta$ for best fiber transmission and for a fiber modal content that is close to single mode as much as possible. In other words, we seek a cladding tubular arrangement, core size and shape such that the quantity $\langle \varphi_{clad} | \Delta n^2 | \varphi_{core} \rangle$ is reduced as much as possible [4]. Consequently, by simply recalling that cladding modes with a large azimuthal-number *m* favors IC guidance, which can be illustrated by the fact that the power overlap between a low-azimuthal number core mode (i.e. HE$_{11}$) and a high azimuthal-number cladding mode scales as $m^{-2}$, it is easy to note that increasing *m* implies thinning the silica tube and increasing its radius. Recently, an empirical study showed that the confinement loss scales as $\alpha_{CL} \propto t/R_t$ [25], which is consistent with the scaling of *m* with $t/D$, with D being the core contour perimeter [11], [12]. However, satisfying these rules is constrained by fabrication limitation and the fiber modal content. Indeed, keeping the circular shape of the tube, and minimizing its surface roughness during the fiber draw sets a limit on how small *t* one can tolerate. Moreover, because $R_t$ is interconnected with the fiber core radius $R_c$, $\delta$ and the number of tubes in the cladding ring *N* through the identity $R_t = (R_c \sin(\pi/N) - \delta/2)(1 - \sin(\pi/N))^{-1}$, its increase can have reverse impact on both the confinement loss and the fiber modal content. In order to find the best trade-off between the above parameters, we start by finding out the optimal $\delta$, then the number *N* and finally the fiber core radius.

Figure 2 summarizes the effect of $\delta$ on both the cladding lattice modal spectrum and on the fiber core. Figure 2(a) shows the evolution with $\delta$ of the electric field magnitude of a representative silica mode. The field is that of EH$_{9,2}$ at wavelength of 700 nm, for $t$ = 400 nm (i.e. $F$ = 1.2), and tube radius of $R_t$ = 7μm. The results show that increasing $\delta$ from 2 to 8 μm results in a decrease of the mode $n_{eff}$ from ~1.00062 to 0.99993, and in a stronger light confinement in the silica tube. This is readily shown in the 1D profile along a gap line joining the center of two adjacent tubes (z-line in Fig. 2(a), bottom panel), and where the field amplitude drops by a factor of more than 10 when $\delta$ is increased from 2 to 8 μm. This trend favors larger inter-tube gap for IC, however this would make sense only if the fiber core mode profile is not altered and kept its zero-order Bessel profile. Figure 2(b) shows the evolution of the HE$_{11}$ core mode of a single-ring tube lattice fiber when the gap is increased. Here, the single-ring cladding is comprised with 8 tubes. The results show that increasing $\delta$ from 2 to 8 μm leads to an increase of the mode-field radius relative to the fiber inner radius by almost 6%. This subsequently entails a stronger optical spatial overlap of the mode with the silica. Furthermore, the bottom panel of Fig. 2(b) clearly shows the increase with $\delta$ of field magnitude along the gap line; indicative of stronger spreading of the core mode in the cladding. Equally, we investigated the effect of $\delta$ on the core mode leakage by calculating the power flux along the fiber radial and transverse directions using the Poynting vector (see Supplement 1). The results clearly show that the transverse power leakage of the HE$_{11}$ core mode at the gaps strongly increases with increasing $\delta$. The leakage increase can reach up to 30 dB when $\delta$ is increased from 2 to 8 μm for λ=530 nm. Consequently, and given the stronger impact of the core radius on the confinement loss ($\alpha_{CL} \propto R_c^{-4}$ [25]) compared to that from the cladding modal change, the optimal $\delta$ will be a trade-off between a value that is sufficiently small to avoid a too high core mode leakage, and sufficiently large to avoid the formation of connecting nodes (which support low azimuthal-number modes) and a too strong coupling between the tubes.

The top of Fig. 3 summarizes the effect of *N* and $\delta$ on the fiber confinement loss. Figure 3(a) shows the CL over a representative spectrum for different *N* and for fibers having $t = 400\,nm$, $R_t = 8\,\mu m$ and $\delta = 5\,\mu m$. The spectrum spans from 480 to 600 nm, and corresponds to *F* range of 1.4-1.75. The loss figure is expressed as $\alpha_{CL} R_c^4$ so to only keep the effect of *N* on the transmission loss trend with increasing *N* regardless of the fiber core radius. The results show that for the same core radius, the CL decreases with *N* increases. However, the decrease rate of the CL with *N* drops as *N* gets larger. For example and choosing a wavelength of 540 nm, $\alpha_{CL} R_c^4$ decreases by a factor of 3 when *N* is changed from 5 to 6. However, it decreases by only less than 20% when the tube number is increased from 8 to 9.

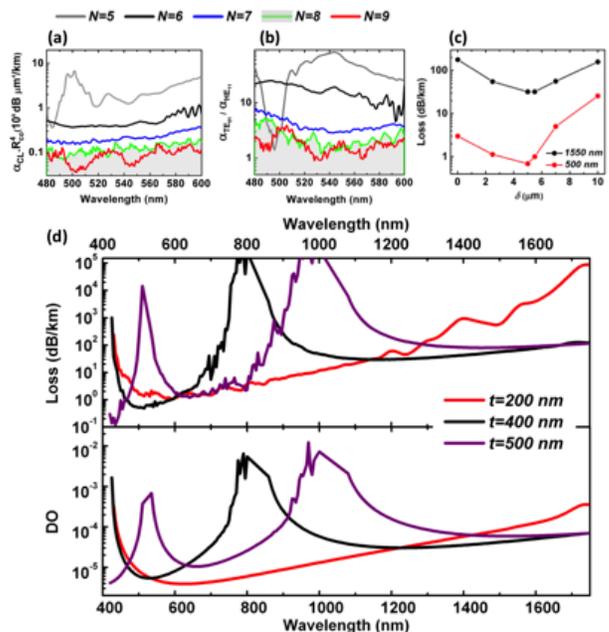

Fig. 3. Four panel showing (a) CL for different *N*, (b) Ratio between TE$_{01}$ and HE$_{11}$ loss, (c) loss spectra versus $\delta$, (d) loss and DO spectra for different *t* values.

This is due to the resonance between the core mode and the air modes of the tubes [17], [19]. On the other hand, the power-ratio of the higher-order core modes $TE_{01}$ and $HE_{11}$, which is an indicator of how well the fiber can operate in a single mode manner, increases as expected with decreasing $N$, thus favoring smaller number of tubes for single mode operation (see Fig. 3(b)). Consequently, and similarly to the finding of the optimal $\delta$, the optimal $N$ is the result of a trade-off between CL-decrease, which scales proportionally with $N$, and the fiber single modedness, which is inversely proportional to $N$, along with bend loss sensitivity which favors smaller fiber core size and thus smaller $N$. We found $N = 8$ to be a good compromise for a sufficiently low CL and a modal content that is dominated by the core fundamental mode $HE_{11}$. Figure 3(c) shows the CL evolution with $\delta$ at two representative wavelengths of a SR-TL HC-PCF with $N = 8$, $t = 400\ nm$, and $R_t = 8\ \mu m$. The results show that the inter-tube gap range of 2-6 μm is suitable for low loss guidance. Based on these design results, we calculated the CL spectra for different tube thickness. Figure 3(d) shows the spectra of CL (top) and the optical overlap of $HE_{11}$ core mode with the cladding dielectric (noted DO) (bottom) at different tube thicknesses ($t$ = 200 nm, 400 nm and 500 nm) for a SR-TL HC-PCF with $N$=8, $R_t = 8\ \mu m$ and $\delta = 2.5\ \mu m$. Within the explored wavelength range of 400-1750 nm, the $t$ = 200 nm fiber exhibits one large transmission band corresponding to the fundamental band with a minimum loss of ~1 dB/km at wavelengths near 600 nm. For the $t$ = 400 nm SR-TL HC-PCF, the spectrum exhibits two transmission bands separated by a high-loss band centered around 800 nm. Here, the lowest loss figure is 0.5 dB/km and occurs at wavelengths near 500 nm in the 1st higher order transmission band. For $t$ = 500 nm SR-TL HC-PCF, we observe a spectrum with the 3 transmission bands, and lowest loss of 0.3 dB/km occurs in the 2nd higher-order band around 480 nm. Below, we explore these results and scaling laws in the fabrication of several fibers with $N$ = 8. The DO spectra show the same resonance structure as CL as expected by IC guidance which relies on the reduction of $\langle \varphi_{clad}|\Delta n^2|\varphi_{core}\rangle$, which itself is proportional to the optical overlap DO. The spectra show that an overlap in the range of $10^{-6}$ is achievable with this type of IC guiding HC-PCF. The minimum obtainable can be reduced by further thinning the tube thickness and operating in the shorter wavelength. This trend results from the fact that the silica cladding modes get more localized with the optical frequency increase. On a note relating to this fiber guidance mechanism mentioned in Supplement 1, it is significant to highlight the wavelength evolution of CL in the fundamental band which cannot be explained with the ARROW picture. For $\lambda >$ 1600 nm, we observe that the CL of the three fibers show different slope with increasing wavelength, and don't scale uniformly with the tube thickness. For example, at $\lambda = 1660$ nm, CL reaches 13480.6 dB/km for $t$ = 200 nm, but is only 94.7 dB/km for $t$ = 400 nm, and 98.4 dB/km for $t$ = 500 nm. In order to explain such a trend, we recall that reducing CL implies reducing $|\langle \varphi_{clad}|\Delta n^2|\varphi_{core}\rangle|^2$. Even though $\langle \varphi_{clad}|\Delta n^2|\varphi_{core}\rangle$ entails expressions of core and cladding field with amplitude and spatial phase that are not separable, we can extract qualitative scaling law by making some approximations. For the case of $HE_{11}$ core mode in the fundamental band (i.e. the radial phase number is $l$=1), the core-cladding field overlap is dominantly in the glass, and if we consider the cladding field amplitude to be constant over the whole glass regions, we can write the approximate proportionality identity $|\langle \varphi_{clad}|\Delta n^2|\varphi_{core}\rangle|^2 \sim m^{-2} DO$. Consequently, the observed CL trend with the tube thickness can be explained by examining DO and the azimuthal number $m$ of the cladding mode in the vicinity of the fiber-core $HE_{11}$ effective index at $\lambda = 1660$ nm (see Supplement 1 for more details). The results show that the corresponding cladding modes are of the family of $EH_{6,1}$ for $t$ = 200 nm fiber, of $EH_{12,1}$ for $t$ = 400 nm, and of $EH_{15,1}$ for $t$ = 500 nm, in consistency with the scaling law $m \propto 1/t$ mentioned above, and their respective $m^{-2} DO$ values follow the CL order as found numerically.

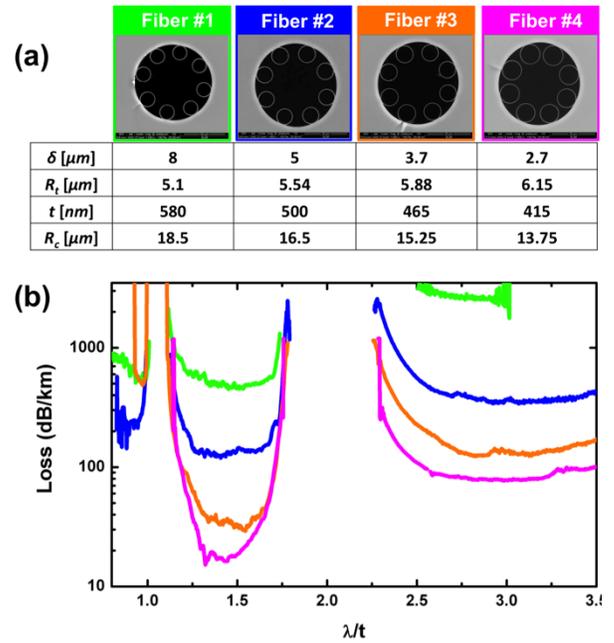

Fig. 4. (a) SEM pictures of 4 fabricated HC tubular lattice fibers with corresponding geometrical parameters ($\delta$, $R_t$, $t$, $R_c$). Experimental evolution of the loss spectrum with the gap between the tubes.

### 3. Experimental results

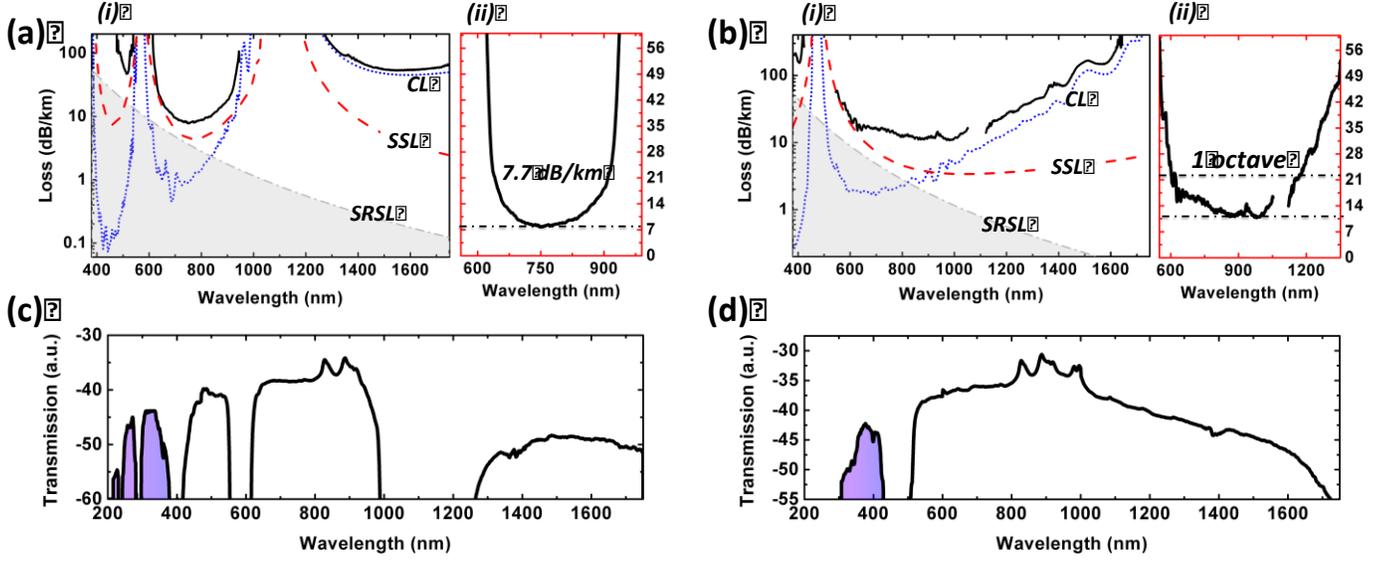

Fig. 5. Top: Measured attenuation spectrum (black curves), calculated CL (dotted blue curves) and SSL (dashed red curves) for Fiber #5 (a(i)) and Fiber #6 (b(i)). Fiber #5 reaches a record of 7.7 dB/km (a(ii)), and Fiber #6 exhibits a loss in the range 10-20 dB/km over one octave (b(ii)). Bottom: Measured transmission of a 2 m-long piece of of Fiber #5 (c) and of Fiber #6 (d) with purple-filled curve highlighting UV/DUV guidance. The blanked data around 1μm in b(i) and b(ii) are due to the supercontinuum stronger power at 1064 nm.

Figure 4(a) shows the micrographs of four fabricated SR-TL HC-PCFs along with their physical properties. Here, we endeavored to vary $\delta$ whilst keeping $R_t$, $t$ and $R_c$ as constant as possible so to find out the optimal value from the stand point of fabrication and transmission performance. The table summarizing these quantities values for the different fibers shows that $\delta$ spans from 2.7 μm (Fiber #4, pinked colored micrograph frame) to 8 μm (Fiber #1, green colored micrograph frame). Within this $\delta$ span, the core radius $R_c$ ranges from 13.75 to 18.5 μm, corresponding to a maximum relative difference between the different fiber core sizes of less than 20%. Similarly, the cladding tubes radius $R_t$ and $t$ range from 5.1 to 6.15 μm and from 415 to 580 nm respectively and the relative difference of both quantities are also less than 20%. Figure 4(b) shows the four fibers measured loss spectra versus $\lambda/t$. As expected over the measured wavelength range of 400-1750 nm and for tube thickness of 415-580 nm, the spectra show three transmission bands corresponding to the fundamental band ($\lambda/t > 2$), 1st order band ($1 < \lambda/t < 2$) and the second-order band ($\lambda/t < 1$). The different fibers loss curves in the fundamental and 1st order bands clearly demonstrate a dramatic reduction of the optical attenuation with $\delta$ decrease; from a near 1 dB/m loss-level for $\delta = 8$ μm (green curve) down to only a few tens of dB/km for $\delta = 2.7$ μm. For the 2nd order band, the trend of the loss-decrease with $\delta$ is observed only for $\delta$ range of 8-5 μm. For smaller inter-tubes gaps, corresponding to tube thicknesses $t$ smaller than 500 nm, the loss increases with $\delta$ decrease (e.g. orange curve compared to the blue curve). Furthermore, a comparison of the rate at which the loss decreases with the gap-shortening shows that the loss in the fundamental band decreases at a larger rate than that in the 1st order band. As a matter of fact it plateaus down to 15 dB/km for $\delta$ range of 3.7-2.7 μm (see pink and orange curves). Neglecting the silica material absorption, a second source of propagation loss is the scattering due to the fiber tubes small scale surface roughness. This stems from frozen capillary-waves during the fiber draw [26] and causes the discrepancy between the measured transmission loss at short wavelengths with the simulated CL, which was calculated for surface-roughness-free fiber. The surface roughness induces both SSL and IC degradation. The wavelength dependence of the SSL for the typical small scale surface irregularities results from the fractional optical overlap with the core-contour $F_{cc}(\lambda)$ and its wavelength scaling of $\lambda^{-3}$ [26]. The SSL expression can be written using the formula $\alpha_{SSL}(\lambda) = \varsigma \cdot F_{cc}(\lambda) \cdot \lambda^{-3}$ [27], with $\varsigma$ being a constant related to the surface roughness root-mean-square height. Consequently, the total propagation loss can be expressed as $\alpha_{tot}(\lambda) = \alpha_{CL}(\lambda) + \alpha_{SSL}(\lambda)$. The effect of the surface roughness on CL in such IC guiding fibers is less obvious, and in order to demonstrate its effects on CL, we considered the simulation of several SR-TL HC-PCF with their tube exhibiting corrugated surfaces to mimic a surface roughness along the azimuthal direction (see Supplement 1 for more details). The results show that a surface with roughness-height as large as 2 nm practically doesn't affect $F_{cc}(\lambda)$. Furthermore, the CL shows little change for wavelength larger than 800 nm, but strongly increases with decreasing wavelengths. This is in consistence with observed trend in the measured spectra of Fig. 4 and with the scaling law $|\langle \varphi_{clad}|\Delta n^2|\varphi_{core}\rangle|^2 \sim m^{-2} DO$, where DO is proportional to $F_{cc}(\lambda)$, and indicate that for the case of small surface roughness, the spatial overlap is much less affected than the modal transverse phase. Finally, this effect of the surface roughness explains the higher transmission loss at short wavelengths in the early IC guiding hypocycloid core-contour HC-PCF [11]–[13], [28], [29].

Now that $\delta$ value impact on the transmission loss of fabricated fibers is demonstrated, we used $\delta \sim 2.5$ μm as a parameter target and undertook two fiber fabrication campaigns with different aims. The first aim consists of the fabrication of fibers with the thinnest tubes possible so to have the broadest fundamental band whilst having loss figures as low as the surface-roughness-induced transmission-loss could permit. The second aim is to have the lowest loss figure possible in the NIR-VIS spectral range. This was undertaken in an iterative process of several fiber draws. As mentioned above, having a too thin cladding tube (typically <250 nm) poses fabrication challenges in keeping the circular shape of the tubes but also increases the surface scattering through larger surface-roughness that results from enhanced surface capillary wave during the draw process. Consequently, reaching the lowest loss in the shortest wavelength will be a compromise between CL (i.e. design limited) and SSL (i.e. fabrication limited). The results of this fabrication campaign are summarized in Fig. 5. The figure shows the loss and transmission spectra of two fibers. The first one (Fiber #5) exhibits an average $\delta$ value of 2.5 μm, $t = 545$ nm and $R_c = 20.5$ μm. Its measured loss spectrum (cut-back between 293 m and 10 m long pieces) shown in Fig. 5(a)(ii) highlights ultra-low loss in the 1st order band with an absolute record transmission-loss for a HC-PCF of 7.7 dB/km at 750 nm. This loss figure is only about a factor 2 larger than the silica Rayleigh scattering limit shown by the grey filled curve.

The second design (Fiber #6) presents thinner tubes ($t$ = 227 nm), thus shifting the fundamental band blue edge to a wavelength as short as 515 nm [Fig. 5(b)(i)]. An ultra-large low-loss window is demonstrated over one octave ranging from 600 to 1200 nm with loss between 10 and 21 dB/km [Fig. 5(b)(ii)]. This is to our knowledge the first time that a HC-PCF combines such a large bandwidth with such a low transmission loss. The measured transmission spectra of a 2 m-long sections from the two fibers are plotted in Figs. 5(c) and 5(d). Both fibers exhibit guidance in the UV domain. Remarkably, Fiber #5 shows three UV transmission-bands spanning down to 220 nm (purple filling color). Reliable measurements of loss spectra for wavelengths shorter than 350 nm were prevented from the limited dynamics of the photospectrometer.

Figures 5(a) and 5(b) also show the theoretical CL and SSL loss spectra. The CL and $F_{cc}(\lambda)$ are calculated numerically for the case of fibers with no surface roughness. The quantity $\varsigma$ is a fit parameter that is determined by interpolating the total loss $\alpha_{tot}(\lambda) = \alpha_{CL}(\lambda) + \alpha_{SSL}(\lambda)$ with the measured transmission loss at the fundamental band. Since at this band both CL and $F(\lambda)$ are not strongly affected by the surface roughness, we can deduce the SSL without the need to measure the surface roughness. The fit was found to better than 20% for both fibers and over the whole fundamental transmission band (see Supplement 1).

The curves of $\alpha_{CL}(\lambda)$ and $\alpha_{SSL}(\lambda)$ show that for Fiber #5, the measured loss is dominated by the CL in the range of 1300-1750 nm. Here, the CL is in the range of 80-45 dB/km, whilst the scattering loss is well below 20 dB/km. However, for the 1st and 2nd higher order bands of the Fiber #5, the SSL is the dominating source of the transmission loss. The CL reaches a minimum near 1 dB/km for the 1st band and 0.1 dB/km in the 2nd band, whilst the SSL level is above 4.4 dB/km for the 1st band, and above 7.4 dB/km for the 2nd band. It is noteworthy that the measured loss at these higher-order bands is found to be higher than the SSL limit, and this discrepancy increases as the wavelength shortens. This is due to the fact that the CL we are considering is calculated for surface-roughness-free fibers. Indeed, introducing a surface roughness increases the CL for short wavelengths and the effect becomes stronger when the wavelengths get shorter (see Supplement 1) in consistency with the measured loss spectra. This trend is also observed for Fiber #6 where the CL dominates for wavelengths larger than 1000 nm, and the surface roughness induced SSL and CL increase becomes the dominant transmission loss for shorter wavelengths (typically less than 800 nm). In addition to the above low transmission loss, these fibers present a low bending sensitivity with for example, 0.03 dB/turn for a 30 cm bend diameter at 750 nm for Fiber #5 (see Supplement 1). The modal content has also been characterized using Spectral and Spatial (S2) imaging technique [30], [31]. Figure 6(a) shows the typical S2 modal content and the evolution with the group delay of the interference signal Fourier transform (GDFT) for fiber-length of L = 5 m and L = 15 m from Fiber #6. The S2 imaging system comprises an InGaAs camera and a tunable laser source with a tuning range of 1010-1070 nm, and a minimum step size of 40 pm. At the output of the fiber, a telescope is used to collect the light recorded through the camera, which is triggered directly from the laser and controlled by a PC. For 5 m long fiber, the LP$_{11}$ mode presents a multi-path interference (MPI) of 21.4 dB which corresponds to a quasi-single mode operation and increases to 23.9 dB for a length of 15 m.

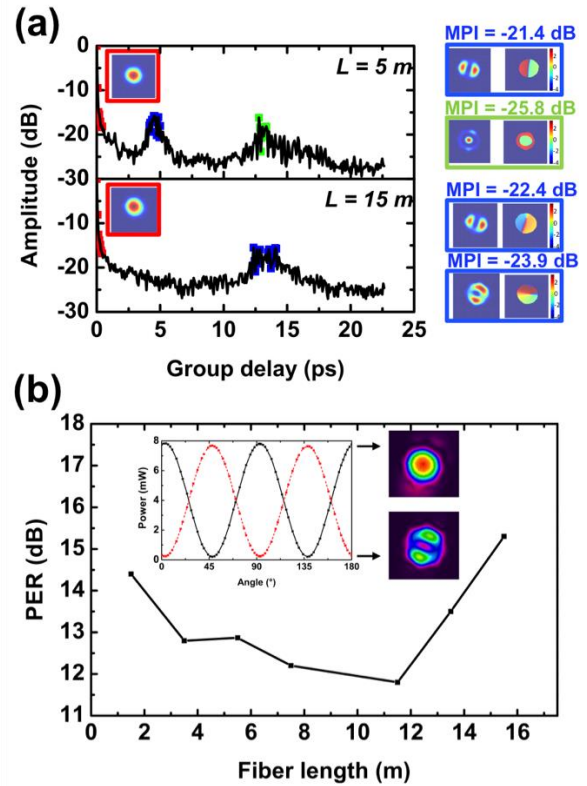

Fig. 6. (a) Group delay curves of the HOM content propagating in the Fiber #6 for 5 m and 15 m long pieces. The reconstructed mode field profiles are indicated with the corresponding MPI. (b) Evolution of the PER versus the fiber length.

The fibers polarization maintaining properties where explored by launching into the fiber a laser beam from linearly polarized laser emitting at 1030 nm and a half-wave plate for polarization control. The transmitted beam is then passed through a polarizing beam-splitter (PBS) and each of its two output beams is recorded by a power-meter and a camera imaging its reconstructed near-field beam profile. Figure 6(b) shows the evolution of the polarization extinction ratio (PER) with the fiber length (Fiber #6 is used). The inset shows the fiber transmitted laser power from each of the 2 output ports of the PBS in function with the half-wave plate angle. The deduced PER was recorded for different fiber lengths. The results show a maximum PER of 15.5 dB achieved with a fiber length of 16 m, and an evolution with the fiber length which comprises a decrease in PER with the fiber length for length shorter than 12 m and then an enhancement in PER as the fiber length increases for length longer than 12 m. It is noteworthy that the profiles of the crossed-polarized beams show that for an input polarization corresponding to a maximum PER, the mode of the dominant polarization orientation is mainly that of HE$_{11}$ whilst the one with the crossed polarization shows a profile of that of the LP$_{11}$ family (Inset of Fig. 6(b)). We believe that this PER evolution with the fiber length and the fact that the mode of the crossed-polarized is that of LP$_{11}$ results from the back-coupling to the core of the light scattering off the core inner surface. Indeed, by virtue of stronger optical overlap of the higher-order core modes with the core-contour, the light that is coupled back from the core-surround surface is likely to couple to higher order modes rather than the HE$_{11}$.

## 4. Conclusion and discussion

In conclusion we fabricated several SR-TL HC-PCFs guiding in the NIR-VIS-UV spectral range and showing ultra-loss. The fiber cladding physical properties were adjusted for optimum confinement loss and close-to-unity modal content using the IC formalism. Among the fabricated fibers we listed one fiber with a record loss of 7.7 dB/km at 750 nm, and a second SR-TL HC-

PCF that combines an octave-wide transmission band with transmission loss in the range of 10-20 dB/km. We have shown that the transmission loss is limited by CL for wavelength longer than ~1 µm, indicating that improving the loss at these spectral ranges and with core diameters comparable to the ones reported here would require a different cladding design with stronger IC. Nested tube lattice HC-PCFs [32], [33] are excellent candidates for this aim as their cladding modes exhibit larger azimuthal number than the reported tubular lattice for the same $n_{eff} - \omega$. The challenge with such nested tubular lattice is the difficulty in controlling the different tube thicknesses during the fabrication process and to keep them the same and constant throughout the draw. On the other hand, we have shown that for shorter wavelengths, typically less than 800 nm, the fiber transmission performance is no longer limited by the fiber design but the surface roughness. The latter affects the propagation loss of SR-TL HC-PCFs via both SSL and CL. The surface roughness induced CL increase is due to the phase-mismatch degradation between the core and the cladding modes. Improving the transmission loss for wavelength shorter than 800 nm would require reducing the surface roughness during the draw, and the cladding lattice design of any hypocycloid core-contour HC-PCF will play no significant role until the transmission loss reaches the level of 1 dB/km or below. The reported results represent an important tool in improving the loss in IC guiding HC-PCF and could revive the prospect of developing optical fibers with transmission loss much below the silica Rayleigh scattering limit.

**Funding**. The authors acknowledge funding from the European Hiperdias project, ANR PHOTOSYNTH, Σ_LIM Labex Chaire, Maturation, region Limousin and AFSOR.

**Acknowledgment**. The authors thank the PLATINOM platform for the help in the fiber fabrication.

See Supplement 1 for supporting content.

## REFERENCES

[1] T. A. Birks, P. J. Roberts, P. S. J. Russell, D. M. Atkin, and T. J. Shepherd, "Full 2-D photonic bandgaps in silica/air structures," *Electron. Lett.*, vol. 31, pp. 1941–1943, 1995.

[2] F. Benabid and P. J. Roberts, "Linear and nonlinear optical properties of hollow core photonic crystal fiber," *J. Mod. Opt.*, vol. 58, no. 2, pp. 87–124, Jan. 2011.

[3] P. S. J. Russell, P. Holzer, W. Chang, A. Abdolvand, and J. C. Travers, "Hollow-core photonic crystal fibres for gas-based nonlinear optics," *Nat Phot.*, vol. 8, no. 4, pp. 278–286, Apr. 2014.

[4] F. Couny, F. Benabid, P. J. Roberts, P. S. Light, and M. G. Raymer, "Generation and photonic guidance of multi-octave optical-frequency combs.," *Science*, vol. 318, no. 5853, pp. 1118–21, Nov. 2007.

[5] J. von Neumann and E. Wigner, "Über merkwürdige diskrete Eigenwerte," *Phys. Z.*, vol. 30, p. 465–467, 1929.

[6] C. W. Hsu, B. Zhen, A. D. Stone, J. D. Joannopoulos, and M. Soljačić, "Bound states in the continuum," *Nat. Rev. Mater.*, vol. 1, p. 16048, Jul. 2016.

[7] C. W. Hsu, B. Zhen, J. Lee, S.-L. Chua, S. G. Johnson, J. D. Joannopoulos, and M. Soljacic, "Observation of trapped light within the radiation continuum," *Nature*, vol. 499, no. 7457, pp. 188–191, Jul. 2013.

[8] H. Friedrich and D. Wintgen, "Interfering resonances and bound states in the continuum," *Phys. Rev. A*, vol. 32, no. 6, pp. 3231–3242, 1985.

[9] J. M. Zhang, D. Braak, and M. Kollar, "Bound States in the Continuum Realized in the One-Dimensional Two-Particle Hubbard Model with an Impurity," *Phys. Rev. Lett.*, vol. 109, no. 11, p. 116405, Sep. 2012.

[10] F. Couny, F. Benabid, and P. S. Light, "Large-pitch kagome-structured hollow-core photonic crystal fiber," *Opt. Lett.*, vol. 31, no. 24, p. 3574, Dec. 2006.

[11] Y. Wang, F. Couny, P. J. Roberts, and F. Benabid, "Low Loss Broadband Transmission In Optimized Core-shape Kagome Hollow-core PCF," in *Conference on Lasers and Electro-Optics 2010*, 2010, p. CPDB4.

[12] Y. Y. Wang, N. V Wheeler, F. Couny, P. J. Roberts, and F. Benabid, "Low loss broadband transmission in hypocycloid-core Kagome hollow-core photonic crystal fiber.," *Opt. Lett.*, vol. 36, no. 5, pp. 669–71, Mar. 2011.

[13] B. Debord, M. Alharbi, T. Bradley, C. Fourcade-Dutin, Y. Y. Wang, L. Vincetti, F. Gérôme, and F. Benabid, "Hypocycloid-shaped hollow-core photonic crystal fiber Part I: arc curvature effect on confinement loss.," *Opt. Express*, vol. 21, no. 23, pp. 28597–608, Nov. 2013.

[14] B. Debord, M. Alharbi, L. Vincetti, A. Husakou, C. Fourcade-Dutin, C. Hoenninger, E. Mottay, F. Gérôme, and F. Benabid, "Multi-meter fiber-delivery and pulse self-compression of milli-Joule femtosecond laser and fiber-aided laser-micromachining.," *Opt. Express*, vol. 22, no. 9, pp. 10735–46, May 2014.

[15] A. D. Pryamikov, A. S. Biriukov, A. F. Kosolapov, V. G. Plotnichenko, S. L. Semjonov, and E. M. Dianov, "Demonstration of a waveguide regime for a silica hollow - core microstructured optical fiber with a negative curvature of the core boundary in the spectral region > 3.5 µm," *Opt. Express*, vol. 19, no. 2, pp. 1441–1448, Jan. 2011.

[16] F. Yu, W. J. Wadsworth, and J. C. Knight, "Low loss silica hollow core fibers for 3-4 µm spectral region.," *Opt. Express*, vol. 20, no. 10, pp. 11153–8, May 2012.

[17] T. D. Bradley, Y. Wang, M. Alharbi, B. Debord, C. Fourcade-Dutin, B. Beaudou, F. Gerome, and F. Benabid, "Optical Properties of Low Loss (70dB/km) Hypocycloid-CoreKagome Hollow Core Photonic Crystal Fiber for Rb and Cs BasedOptical Applications," *J. Light. Technol.*, vol. 31, no. 16, pp. 2752–2755, Aug. 2013.

[18] F. Benabid, F. Gerome, B. Debord, and M. Alharbi, "Kagome PC fiber goes to extremes for ultrashort-pulse lasers," *Laser Focus World*, vol. 50, no. 9, pp. 29–34, 2014.

[19] L. Vincetti and V. Setti, "Waveguiding mechanism in tube lattice fibers.," *Opt. Express*, vol. 18, no. 22, pp. 23133–46, Oct. 2010.

[20] M. Michieletto, J. K. Lyngsø, C. Jakobsen, J. Lægsgaard, O. Bang, and T. T. Alkeskjold, "Hollow-core fibers for high power pulse delivery," *Opt. Express*, vol. 24, no. 7, pp. 7103–7119, 2016.

[21] N. M. Litchinitser, a K. Abeeluck, C. Headley, and B. J. Eggleton, "Antiresonant reflecting photonic crystal optical waveguides.," *Opt. Lett.*, vol. 27, no. 18, pp. 1592–4, Sep. 2002.

[22] We recall as established in solid-state physics that the periodicity is not conditional to the existence of bandgap.

[23] M. M. Z. Kharadly and J. E. Lewis, "Properties of dielectric-tube waveguides," *Electrical Engineers, Proceedings of the Institution of*, vol. 116, no. 2. pp. 214–224, 1969.

[24] M. A. Duguay, Y. Kokubun, T. L. Koch, and L. Pfeiffer, "Antiresonant reflecting optical waveguides in SiO2-Si multilayer structures," *Appl. Phys. Lett.*, vol. 49, no. 1, 1986.

[25] L. Vincetti, "Empirical formulas for calculating loss in hollow core tube lattice fibers," *Opt. Express*, vol. 24, no. 10, pp. 10313–10325, 2016.

[26] P. Roberts, F. Couny, H. Sabert, B. Mangan, D. Williams, L. Farr, M. Mason, a Tomlinson, T. Birks, J. Knight, and P. St J Russell, "Ultimate low loss of hollow-core photonic crystal fibres." *Opt. Express*, vol. 13, no. 1, pp. 236–44, Jan. 2005.

[27] F. Poletti, "Nested antiresonant nodeless hollow core fiber," *Opt. Express*, vol. 22, no. 20, p. 23807, Oct. 2014.

[28] T. D. Bradley, Y. Y. Wang, M. Alharbi, and B. Debord, "Hypocycloid-core Kagome hollow core photonic crystal fiber for Rb and Cs based optical applications," *J. Light. Technol.*, vol. 31, no. 16, pp. 1–4, 2013.

[29] M. Alharbi, T. Bradley, B. Debord, C. Fourcade-Dutin, D. Ghosh, L. Vincetti, F. Gérôme, and F. Benabid, "Hypocycloid-shaped hollow-core photonic crystal fiber Part II: cladding effect on confinement and bend loss." *Opt. Express*, vol. 21, no. 23, pp. 28609–16, Nov. 2013.

[30] J. W. Nicholson, A. D. Yablon, S. Ramachandran, and S. Ghalmi, "Spatially and spectrally resolved imaging of modal content in large-mode-area fibers," *Opt. Express*, vol. 16, no. 10, pp. 7233–7243, 2008.

[31] P. Calvet, "Mise en forme spataile dans une fibre optique microstructurée pour la réalisation d'amplificateurs lasers tout fibrés pour les pilotes des lasers de puissance."

[32] W. Belardi and J. C. Knight, "Hollow antiresonant fibers with reduced attenuation," *Opt. Lett.*, vol. 39, no. 7, pp. 1853–1856, 2014.

[33] M. S. Habib, O. Bang, and M. Bache, "Low-loss hollow-core silica fibers with adjacent nested anti-resonant tubes," *Opt. Express*, vol. 23, no. 13, pp. 17394–17406, 2015.

# Ultra-low transmission loss (7.7 dB/km at 750 nm) inhibited-coupling guiding hollow-core photonic crystal fibers with a single ring of tubular lattice: supplementary material

This document provides supplementary information to "Ultra-low transmission loss (7.7 dB/km @750 nm) inhibited-coupling guiding hollow-core photonic crystal fibers with a single ring of tubular lattice". The supplementary material is organized as follows. In §1 we compare the impact of the tubular lattice cladding geometry arrangement on its modal spectrum by studying the density of photonic states (DOPS). In §2, we recall some properties of optical guidance by PBG and IC and how the ARROW picture is related to them. §3 shows the impact of the inter-tube gap on the fiber core $HE_{11}$ mode transverse leakage by examining the Poynting vector. In §4 we show the cladding mode structure for long wavelengths in the fundamental band of three fibers with different thickness, and highlight the role of the mode azimuthal numbers on the confinement loss. In §5 we present the effect of the fiber core-contour transverse fluctuation on the confinement loss. §6 shows the fit of a measured loss spectrum with the total theoretical transmission loss. Finally, in §7 we report on experimental results on bend loss sensitivity of two fabricated fibers.

## 1. DOPS VERSUS THE CLADDING DESIGN

This section aims to demonstrate that under the large pitch regime (*i.e.* $\Lambda \gg \lambda$) a photonic structure consisting of a lattice of waveguiding dielectric materials, exhibits a dispersion diagram (*i.e.* modal spectrum or density of photonic states) that depends very little on the lattice pitch or on its geometrical arrangement. Here, we consider the tubular lattice of glass tubes and investigate the impact of changing the cladding arrangement on its modal spectrum. Calculations of DOPS have been carried out for both triangular and square lattice and for three different $\boldsymbol{\delta}$ = 0.04 μm, 0.5 μm, 2.5 μm. The tube thickness and radius were constant to be $t$ = 400 nm and $R_t$ = 7 μm. The pitch is given by $\boldsymbol{\Lambda = 2R_t + \delta}$. Figure 1 shows six DOPS diagrams corresponding to the different lattice parameters mentioned above. Within $n_{eff} - F$ range explored here, all the DOPS diagrams show a quasi-continuum of modes similar to the one shown in the main text. Irrespectively of the lattice geometry or the lattice pitch, the DOPS diagrams show similar overall landscape with very little variation between them. Figure 2 further corroborates this fact by showing the mode profiles at $F$ = 0.8. These modes show the same transverse structure with the same azimuthal and radial numbers. The only difference is in their $n_{eff}$ which exhibit a relative difference value of less than 0.01%. These results, which show very weak effect of the lattice pitch and geometry of the DOPS, are attributed to the fact that the modes are highly localized with little interaction with their close neighboring guiding sites. This fact has been first observed in PCF by Litchinitser *et al.* [1] and then explained using the photonic analogue of tight binding model by Couny *et al.* [2] in the context of investigating how photonic bandgap form in PBG guiding HC-PCF. Consequently, the optical fiber formed by introducing a core defect in a tubular lattice in a large pitch regime won't strongly depend on the lattice pitch nor on its geometrical symmetry.

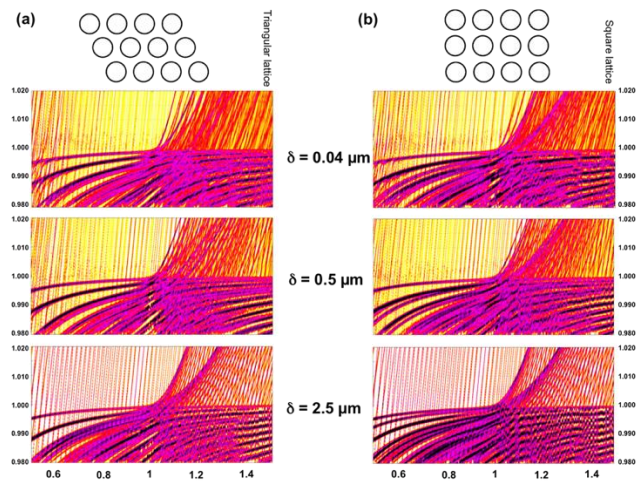

**Fig. 1**. DOPS diagram for (a) triangular and (b) square lattice with δ = 0.04 μm, 0.5 μm and 2.5 μm.

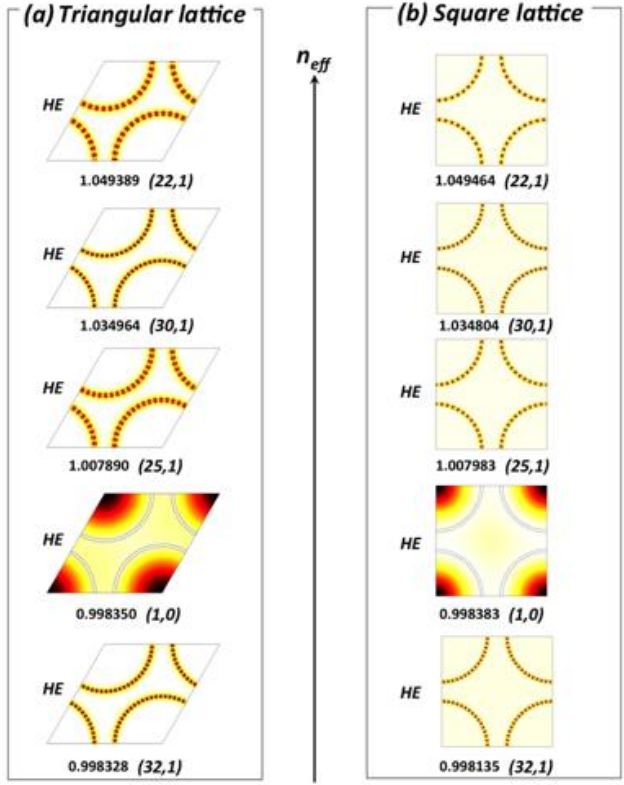

**Fig. 2.** Evolution of the transverse electric field amplitude of the modes at $F = 0.8$ in the case of (a) triangular and (b) square lattice cladding with $\delta = 2.5$ μm.

## 2. Notes on the distinctive features of ARROW, PBG and IC

Given, the recent surge in the use of ARROW terminology, especially within the framework of hypocycloid core-contour hollow-core fibers (also sometimes called negative curvature hollow fibers, or anti-resonant hollow fibers) and the fact that this ARROW picture has been used invariably to describe PBG guiding fibers [3] and IC guiding fibers [4], we deem it important to provide some clarifications and distinctive properties so as to distinguish between PBG, IC and ARROW.

We start by recalling that the term of anti-resonant reflection was initially commonly-used in interferometry to designate a Fabry-Perot cavity configuration when its reflectivity is maximum. Here, the reflection is enhanced compared to the Fresnel reflection by destructive interference of the optical wave inside the Fabry-Perot cavity. This effect has been extended to guided optics as early as 1986 by Dugay *et al.* [5] who coined the resulting waveguide ARROW. Here, a planar waveguide with a low-index core layer is surrounded by higher-index thin dielectric layers to provide a transverse optical confinement in the core through the enhanced thin cladding-layers' reflectivity at the anti-resonant wavelength ranges. This ARROW picture was then used to derive the confinement loss of a hollow-core fiber whose cladding is comprised by one or several concentric anti-resonant dielectric rings [6]. Among the findings of this work is an analytical expression of the $HE_{11}$ mode minimum loss of a single anti-resonant hollow fiber for the case of $R_c \gg \lambda$, which is given by:

$$\alpha_{min} = \alpha_{HD}(2.405\,\lambda)\left(2\pi\,R_c\sqrt{n_g^2 - 1}\right)^{-1}$$
$$= \frac{1}{2}(2\pi)^{-3}(2.405)^3\left(n_g^2 - 1\right)^{-1}\left(n_g^2 + 1\right)\lambda^3 R_c^{-4}. \quad (1)$$

Here, $\alpha_{HD}$ is the $HE_{11}$ mode loss in a hollow dielectric fiber (*i.e.* tube thickness is considered infinitely large) deduced by Marcatili and Schmeltzer in the sixties [7].

The $HE_{11}$ minimum loss coefficient expression of equation (1) was written in this form to highlight the loss-decrease in an anti-resonant ring compared to a hollow dielectric fiber. It shows that the loss-decrease factor scales as $\lambda/R_c$.

Furthermore, using this expression for a core diameter of 15 μm, which is equal to that of the first reported Kagome HC-PCF in 2002 [8], one finds a loss range of 7-365 dB/m at 400-1500 nm spectral range. These loss figures were much higher than the measured loss in the aforementioned Kagome fiber, which was less than 3 dB/m over the same wavelength range, thus leading the authors of ref. [8] to rule out the ARROW model as a possible comprehensive explanation for Kagome HC-PCF guidance. Nevertheless, the ARROW terminology was later found use to describe a class of PCF by Litchinitser *et al.* [1]. The authors considered a PCF whose cladding entails isolated high-index inclusions in a lower index matrix, and coined the fiber ARROW PCF because for sufficiently large pitch relative to the high index inclusion diameter and to the optical wavelength, they found that the band edges of the fiber transmission spectrum depend very little on the pitch, and are strongly related to the optical feature of an individual inclusion. This PCF was then proved experimentally and theoretically by Argyros *et al.* [3], [9] to guide via PBG (*i.e.* the DOPS diagram of an infinite cladding doesn't support any propagating mode in the $n_{eff} - \omega$ space region of interest). Furthermore, it has been demonstrated later that the property of the non-dependence of the transmission band edges with the pitch can be explained by the photonic analogue of tight binding model [2], where for sufficiently large values of the product between the lattice pitch and the optical frequency (*i.e.* normalized frequency), the allowed photonic bands don't broaden from a single dispersion curve of an individual inclusion. In analogy with the strong tight-binding regime in solid-state physics, the large pitch regime corresponds here to a regime where the high-index material cladding modes are highly localized and don't strongly interact with their neighboring photonic sites. Thus, these modes are better represented by the maximally localized Wannier functions instead of extended overlapping Bloch states [10], and the DOPS diagram structure is close to that of dispersion curves from a single high index inclusion instead of photonic allowed bands. As a conclusion, the ARROW PCF in reference [1] is an example of a PBG guiding PCF in the large pitch regime. Also, similarly to amorphous materials which can exhibit electronic bandgap, it is noteworthy to stress that the periodicity of a photonic structure is neither a necessary condition nor a defining feature for the existence of PBG.

In parallel, ARROW has also been extensively used very recently as a model to describe IC guiding fibers such as Kagome-lattice fibers or the present tubular lattice fibers. However, none of this recent reported literature on anti-resonant hollow fibers did justify the use of the ARROW terminology for such fibers or described their cladding modal spectrum, and seems to settle for the fact that the transmission band edges are located at $\lambda_l = (2\,t/l)\sqrt{n_g^2 - 1}$ being a sufficient and a defining feature for ARROW guidance. This feature or the reflection enhancement from thin dielectric cannot explain why fibers whose cladding have the same structure and the same glass thickness exhibit a dramatic difference in confinement loss depending on whether the fiber core-contour is hexagonal, circular or hypocycloid shape, and this is so despite presenting comparable core diameters [11], [12]. This is expected because the original proposals of ARROW [5] don't treat the cladding modal structure and how they couple or don't couple with the core mode in the anti-resonant bands. Instead, the model explains the core mode confinement through the cladding surface reflection enhancement which is in contrast with the much lower loss (<10 dB/km) observed in hypocycloid core-contour HC-PCF. Furthermore, one example that shows that the guidance in these optical fibers cannot be reduced to the difference in glass thickness, and highlights the role of the optical overlap and the azimuthal phase of the cladding modes is the CL behavior at the longer wavelength of the fundamental band (see section 2 in the main text and section 4 of the present supplement). All this is unforeseen by the ARROW picture because it completely neglects the complexity of the cladding modal structure. In fact in the original proposals of ARROW [5, 6] only claddings with just one transverse coordinate dependence (*i.e.* multilayered structures in rectangular or cylindrical coordinates) and only the radial phase matching between core and cladding modes are considered. In fibers with a more complex cladding structure and core-cladding interface such as Kagome or tubular lattice fibers these approximations make it inadequate.

In conclusion, and using the condensed matter physics notions which were transferred to optics these past decades, we take the problem of guidance in an optical fiber by describing its cladding modal structure and then consider the possible core modes that have no or limited coupling pathways to the cladding modes. Within this framework, we can classify light guidance in optical fibers into two types. The first one relies on the fact that the $n_{eff} - \omega$ of a core defect mode (*i.e.* bound photonic state) lies outside any cladding mode continuum (*i.e.* PBG). In this situation, the core mode has no pathway for it to radiate away through the cladding. This includes total internal reflection guidance where the guided mode effective index is larger than the cladding material index, and thus in $n_{eff} - \omega$ space where the cladding doesn't support any radiating wave (*i.e.* PBG). The second one (IC guidance) corresponds to a situation where the guided mode $n_{eff} - \omega$ lies inside the cladding mode continuum, but its coupling to it can be completely suppressed (*i.e.* BIC) or strongly reduced (*i.e.* QBIC) via symmetry incompatibility [13], and thus preventing or reducing the radiation channels though which a core mode can leak out.

## 3. POYNTING VECTOR STUDY

The fiber confinement loss can be efficiently investigated by considering the imaginary part of the mode eigenvalue [14]. Unfortunately this approach doesn't give any information about the leakage mechanism. Here we propose to investigate the fiber leakage mechanisms by considering the power flux along the fiber radial direction.

The density of power along this direction is given by the radial component of the Poynting vector:

$$p_r = \frac{1}{2}\vec{E} \times \vec{H} \cdot \hat{r}, \quad (2)$$

with $\vec{E}$, and $\vec{H}$ being the electric and magnetic fields of the mode and $\hat{r}$ the radial unit vector. $p_r$ describes the modal power density flowing along the radial direction due to the leaky nature of the modes. The integration of $p_r$ along a circle $l$ surrounding the microstructured region of the fiber gives the leaked power per unit of length:

$$P_r = \oint_l p_r \, dl. \quad (3)$$

Figure 3 compares the confinement loss computed by considering the imaginary part of the mode eigenvalue with that computed by starting from $P_r$ through the formula:

$$CL = P_r/P_m, \quad (4)$$

with $P_m$ the power of the $HE_{11}$ core mode. The figure clearly shows that the imaginary part of the mode eigenvalue and the formula of equation (4) give the same CL.

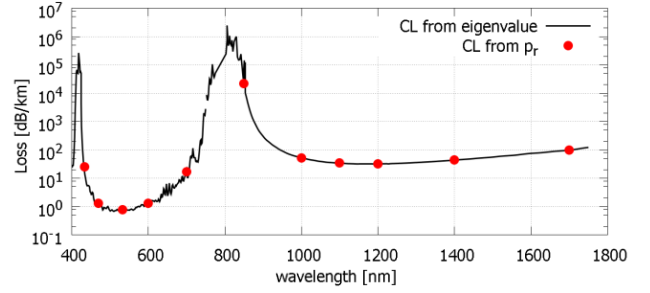

**Fig. 3.** Comparison between CL computed through the mode eigenvalue (solid black line) and through $P_r$ (red points) of a fiber with $\delta = 5.0 \, \mu m$.

to investigate the effect of changing $\delta$ on the mode leakage and to analyze the azimuthal distribution of the confinement loss.

Figure 4 shows the transverse profile of the normalized radial component $p_{r_n}$:

$$p_{r_n} = \frac{p_r}{p_{z_{max}}},$$

being $p_{z_{max}}$ the maximum value of the longitudinal component of the Poynting vector on the fiber cross section. Four fibers having different $\delta$ at two representative wavelengths are considered: 530 nm, and 1200 nm, corresponding to the CL minimum in the fundamental and first transmission band.

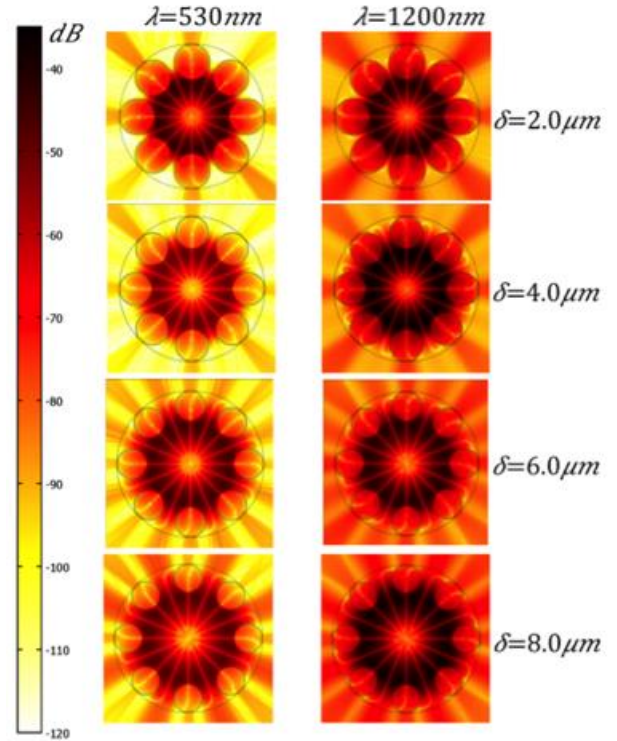

**Fig. 4**. (a) Transverse distribution of $p_{r_n}$ for four fibers with $\delta$ = 2 μm, 4 μm, 6 μm, 8 μm, and at two wavelengths: 530 nm (*lhs* column) and 1200 nm (*rhs* column).

Figure 4 clearly shows that for small inter-tube gap values (e.g $\delta$ = 2.0 μm), the mode power mainly leaks out through the touching regions between the tubes and the surrounding silica, whereas for $\delta$>6.0 μm additional leakage through the inter-tube gaps takes place. This trend is further shown in Fig. 5, which shows the angular distribution of $p_{r_n}$ at a radius of 40 μm. Tube-silica touching points correspond to an angle of 0° and multiples of 45°. The values of the maxima corresponding to the touching point do not significantly depend on $\delta$. Lower values are observed at 530 nm than at 1200 nm, in agreement with CL spectra. Conversely, the values corresponding to the tube-tube gaps dramatically depend on $\delta$. At 530 nm, with $\delta = 2$ μm they are 20 dB lower than the touching points ones, whereas they are 10 dB higher with $\delta = 8$ μm. The dependence is a little bit weaker at 1200 nm, however the variation is about 20 dB.

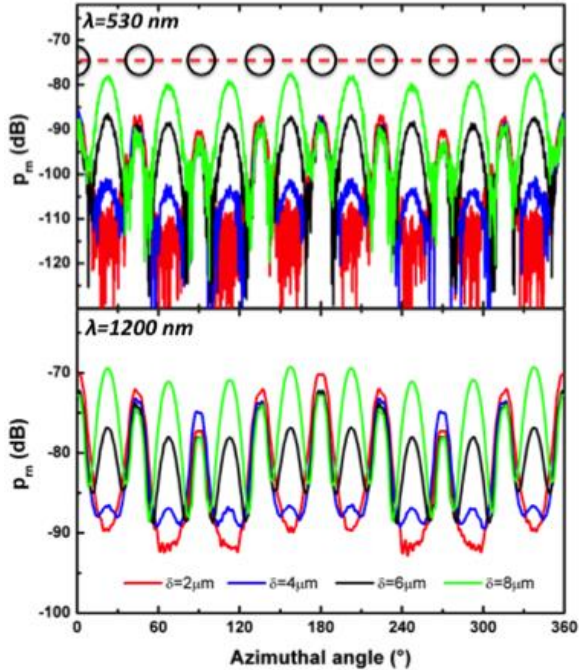

**Fig. 5** Azimuthal distribution of $pr_n$ in the silica surrounding the microstructured cladding. The 8 tubes are placed at 0° and multiple of 45°. The position of the tubes along the considered perimeter (fixed at 40 μm) are reprented by black circles.

## 4. Loss in fundamental band effect of thickness and *m*

Here, we explore how the azimuthal component of the electromagnetic field profile of the cladding modes affects the confinement loss. To this purpose, we examine the cladding modes at a fixed wavelength in the fundamental band (*i.e.* the radial phase index is $l=1$) of three fiber claddings with different thicknesses.

Figure 6 shows the dispersion diagram of a tubular lattice near λ=1660nm with different thickness (dotted curves) and near the effective index of the HE$_{11}$ of the corresponding SR-TL HC-PCF with *N*=8 (green curves), along with the electric field amplitude profile of some representative modes. Figure 6(a) corresponds to a tube lattice with t = 200 nm (*i.e.* F = 0.262), Fig. 6(b) to t = 400 nm (F = 0.525) and Fig. 6(c) to t = 600 nm (*F* = 0.787). These profiles show that the azimuthal number of the mode near the effective index of the HE$_{11}$ fiber core mode at λ = 1660 nm correspond to *m*=6, 12 and 15 for t = 200 nm, 400 nm and 600 nm respectively.

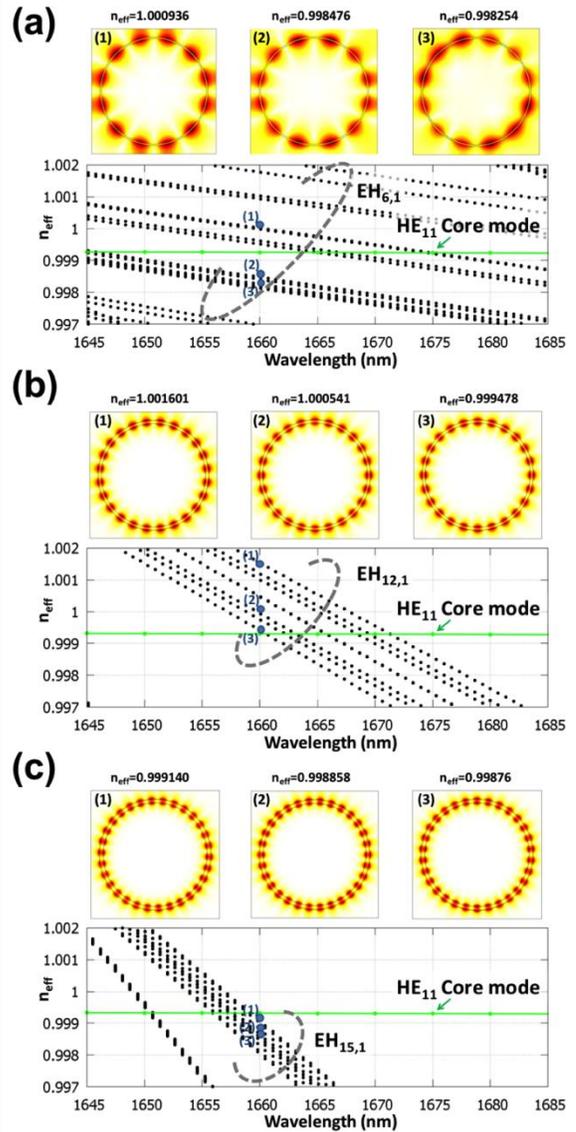

**Fig. 6.** Top: electric field pattern of the cladding modes in resonance with the core mode at λ = 1660 nm corresponding to F = 0.262 for t = 200 nm (a), F = 0.525 for t = 400 nm (b), and F = 0.787 for t = 600 nm (c). They have azimuthal number m = 5, m = 12, m = 16 respectively. Bottom: dispersion curves of the FM for three fibers with t = 200 nm, t = 400 nm, t = 600 nm versus the normalized frequency.

## 5. Effect of surface roughness on CL

In order to explore qualitatively the effect of the tube surface roughness on the CL, we considered a tube whose inner and outer radius exhibit a spatial modulation. The expression of the radius can be written as $\tilde{r}_{out}(\theta) = r_{out} + h\sin(2\pi p\theta)$. Here, $r_{out}$ stands for the outer radius of a tube with no roughness. The quantity $h$ is the roughness height and $p$ is the index of the modulation spatial period. Figure 7(a) shows schematically the fiber and the tube with modulated surfaces. Figure 7(b) shows the spectrum of fractional optical overlap with the cladding, *F*, and the confinement loss, CL, for different spatial periods. Here, the SR-TL HC-PCF is taken to have $R_t = 7$ μm, $\delta = 2$ μm, $t = 400$ nm and $h = 2$ nm. The results show that the effect of the surface roughness on $F$ is negligible for the fundamental and 1st order bands, and start to have a small impact at the second-order band corresponding to wavelengths shorter than 400 nm. Furthermore, the CL spectra show an increase in CL only for higher-order bands (wavelengths shorter than 800 nm) while the fundamental band shows almost no change. The increase in CL gets stronger with decreasing wavelength where the loss is increased by a factor of more than 1000 compared to a fiber with no roughness in the 2nd order band. The observed CL increase in fibers with the corrugated tube surfaces at shorter wavelengths can be inferred by recalling the coupled-mode theory result which shows that a coupling coefficient of a waveguide with a perturbation is inversely proportional to the wavelength. Consequently, and because the IC guidance relies on decreasing the overlap integral between a core mode and the cladding modes, a surface roughness creates a perturbation in the spatial index profile of the silica tube. This perturbation and its periodicity affect the coupling between the cladding modes with the $HE_{11}$ core mode, especially at short wavelengths [15]. This effect partly explains the measured spectra of the different fabricated fibers shown in Fig. 3 of the manuscript.

calculated using the transverse index profile of the fabricated fibers with no surface roughness. The second term corresponds to the surface scattering loss with $\varsigma$ being a fit parameter that is determined by interpolating the total loss $\alpha_{tot}(\lambda)$ with the measured transmission loss at the fundamental band. The bottom of the Fig. 8 shows the experimental loss (black curve) and the total loss $\alpha_{tot}$ (red curve) spectra. The fit was found to be better than 20% for both fibers over the fundamental transmission band (top of Fig. 8). Since at this band both CL and $F_{cc}(\lambda)$ are not strongly affected by the surface roughness (see §5), in high order transmission bands the SSL can be deduced without the need to measure the surface roughness. One can note a maximum difference of 2.5 dB in the 2nd order transmission band and 45 dB in the third one between total and experimental loss.

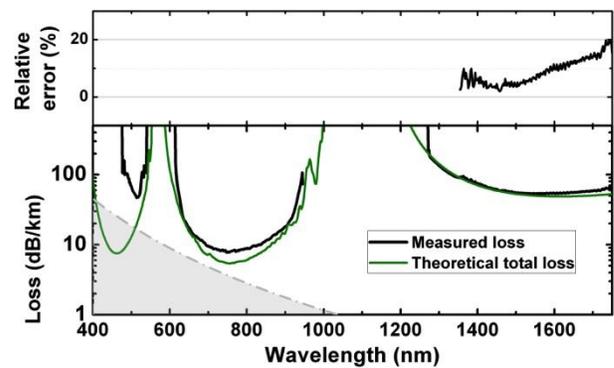

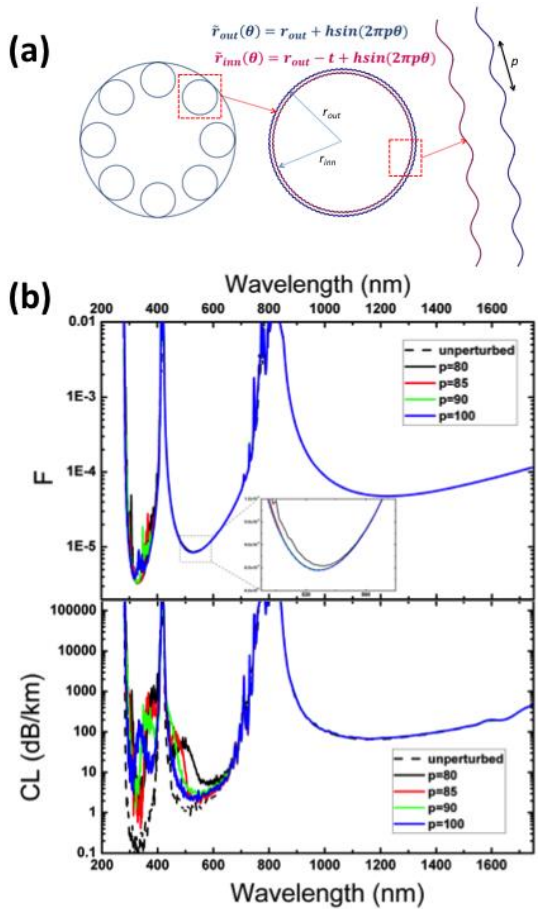

**Fig. 7.** Spectra of fractional optical overlap with the cladding, $F$, and the confinement loss, CL, for different spatial periods $p$ (unperturbed, $p = 80$, $p = 85$, $p = 90$, $p = 100$).

**Fig. 8.** Bottom: experimental (black curve) and theoretical total loss (red curve) spectra. Top: relative error between experimental and theoretical total loss over the fundamental band.

### 7. BENDING LOSS MEASUREMENT

The sensitivity of the both record fibers to bend was investigated. Transmission spectra were measured from the output of the fibers at different bend radii over a 20 m long piece. Typical bend radii (radius of curvature noted $R_c$) used were 20 cm, 15 cm, 7.5 cm, 5 cm and 2.5 cm. As expected, bend losses increase with the decrease in bending radius and with the decrease of the silica core contour thickness. At the wavelength where the minimum attenuation is reached, both fibers show acceptable values respectively of 0.03 dB/turn at 750 nm for Fiber #5 and 0.7 dB/turn at 1000 nm for Fiber #6 when $R_c$ is superior to 15 cm (see Fig. 9).

### 6. Loss spectrum fit

Figures 3(a) and 4(b) of the manuscript shows the measured loss along with the theoretical loss $\alpha_{tot}(\lambda) = \alpha_{CL}(\lambda) + \alpha_{SSL}(\lambda)$ where $\alpha_{SSL}(\lambda) = \varsigma \cdot F_{cc}(\lambda) \cdot \lambda^{-3}$. The first term of the $\alpha_{tot}(\lambda)$, the CL contribution, is numerically

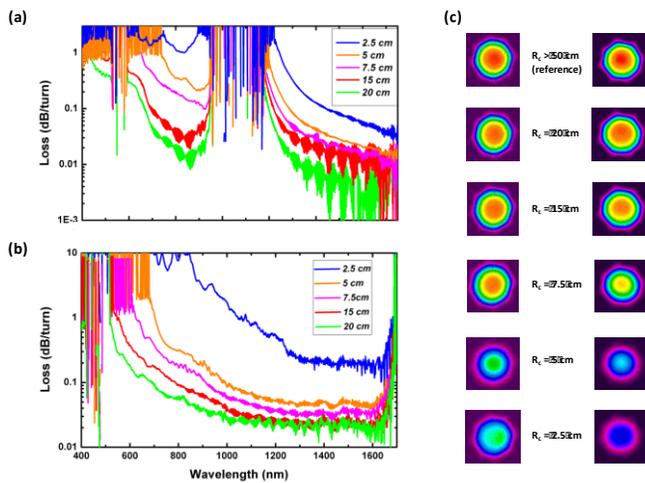

**Fig. 9.** Bend loss spectra measured at different bend radius for (a) Fiber #5 and (b) Fiber #6; (c) Recorded near fields are added for the different bending radius used (left: Fiber #5 / right: Fiber #6).

## References

[1] N. M. Litchinitser, a K. Abeeluck, C. Headley, and B. J. Eggleton, "Antiresonant reflecting photonic crystal optical waveguides.," *Opt. Lett.*, vol. 27, no. 18, pp. 1592–4, Sep. 2002.

[2] F. Couny, F. Benabid, P. J. Roberts, M. T. Burnett, and S. A. Maier, "Identification of Bloch-modes in hollow-core photonic crystal fiber cladding," *Opt. Express*, vol. 15, no. 2, pp. 325–338, 2007.

[3] A. Argyros, T. A. Birks, S. G. Leon-Saval, C. M. B. Cordeiro, F. Luan, and P. S. J. Russell, "Photonic bandgap with an index step of one percent," *Opt. Express*, vol. 13, no. 1, pp. 309–314, Jan. 2005.

[4] F. Couny, F. Benabid, P. J. Roberts, P. S. Light, and M. G. Raymer, "Generation and photonic guidance of multi-octave optical-frequency combs.," *Science*, vol. 318, no. 5853, pp. 1118–21, Nov. 2007.

[5] M. A. Duguay, Y. Kokubun, T. L. Koch, and L. Pfeiffer, "Antiresonant reflecting optical waveguides in SiO2-Si multilayer structures," *Appl. Phys. Lett.*, vol. 49, no. 1, 1986.

[6] J. L. Archambault, R. J. Black, S. Lacroix, and J. Bures, "Loss calculations for antiresonant waveguides," *Journal of Lightwave Technology*, vol. 11, no. 3. pp. 416–423, 1993.

[7] E. A. J. Marcatili and R. A. Schmeltzer, "Hollow Metallic and Dielectric Waveguides for Long Distance Optical Transmission and Lasers," *Bell Syst. Tech. J.*, vol. 43, no. 4, pp. 1783–1809, 1964.

[8] F. Benabid, J. C. Knight, G. Antonopoulos, and P. S. J. Russell, "Stimulated Raman scattering in hydrogen-filled hollow-core photonic crystal fiber.," *Science*, vol. 298, no. 5592, pp. 399–402, Oct. 2002.

[9] A. Argyros, T. A. Birks, S. G. Leon-Saval, C. M. B. Cordeiro, and P. S. J. Russell, "Guidance properties of low-contrast photonic bandgap fibres," *Opt. Express*, vol. 13, no. 7, pp. 2503–2511, 2005.

[10] N. Marzari, A. A. Mostofi, J. R. Yates, I. Souza, and D. Vanderbilt, "Maximally localized Wannier functions: Theory and applications," *Rev. Mod. Phys.*, vol. 84, no. 4, pp. 1419–1475, Oct. 2012.

[11] B. Debord, M. Alharbi, T. Bradley, C. Fourcade-Dutin, Y. Y. Wang, L. Vincetti, F. Gérôme, and F. Benabid, "Hypocycloid-shaped hollow-core photonic crystal fiber Part I: arc curvature effect on confinement loss.," *Opt. Express*, vol. 21, no. 23, pp. 28597–608, Nov. 2013.

[12] M. Alharbi, T. Bradley, B. Debord, C. Fourcade-Dutin, D. Ghosh, L. Vincetti, F. Gérôme, and F. Benabid, "Hypocycloid-shaped hollow-core photonic crystal fiber Part II: cladding effect on confinement and bend loss.," *Opt. Express*, vol. 21, no. 23, pp. 28609–16, Nov. 2013.

[13] C. W. Hsu, B. Zhen, A. D. Stone, J. D. Joannopoulos, and M. Soljačić, "Bound states in the continuum," *Nat. Rev. Mater.*, vol. 1, p. 16048, Jul. 2016.

[14] D. Ferrarini, L. Vincetti, M. Zoboli, A. Cucinotta, and S. Selleri, "Leakage properties of photonic crystal fibers," *Opt. Express*, vol. 10, no. 23, pp. 1314–1319, Nov. 2002.

[15] L. Vincetti and V. Setti, "Extra loss due to Fano resonances in inhibited coupling fibers based on a lattice of tubes.," *Opt. Express*, vol. 20, no. 13, pp. 14350–61, Jun. 2012.